\documentclass[aps, prd, 10pt, twocolumn, superscriptaddress, noshowpacs, preprintnumbers, longbibliography, groupedaddress, bibnotes,hyperref,nofootinbib]{revtex4-1}

\usepackage{amsmath}
\usepackage{amsfonts}
\usepackage{amssymb}
\usepackage{bbold}
\usepackage{epsfig}
\usepackage{graphicx}
\usepackage{bm}
\usepackage{array}
\usepackage{hyperref}
\usepackage{listings}
\usepackage{color}
\usepackage{float}
\usepackage[normalem]{ulem}

\newcommand{\cemit}{\mathcal{C}_{\textrm{emission}}}
\newcommand{\cabsorb}{\mathcal{C}_{\textrm{absorb}}}
\newcommand{\cdirch}{\mathcal{C}_{\textrm{dir-ch}}}

\usepackage{tikz,xcolor,hyperref}

\definecolor{lime}{HTML}{A6CE39}
\DeclareRobustCommand{\orcidicon}{\hspace{-1mm}
	\begin{tikzpicture}
	\draw[lime, fill=lime] (0,0) 
	circle [radius=0.16] 
	node[white] {{\fontfamily{qag}\selectfont \tiny \,ID}};
	\draw[white, fill=white] (-0.0525,0.095) 
	circle [radius=0.007];
	\end{tikzpicture}
	\hspace{-3mm}
}

\foreach \x in {A, ..., Z}{\expandafter\xdef\csname orcid\x\endcsname{\noexpand\href{https://orcid.org/\csname orcidauthor\x\endcsname}
			{\noexpand\orcidicon}}
}

\begin{document}

\title{Neutrino Decoupling  Is Altered by Flavor Conversion}

\author{Shashank Shalgar\orcidA{}}
\affiliation{Niels Bohr International Academy \& DARK, Niels Bohr Institute,\\University of Copenhagen, Blegdamsvej 17, 2100 Copenhagen, Denmark}

\author{Irene Tamborra\orcidB{}}
\affiliation{Niels Bohr International Academy \& DARK, Niels Bohr Institute,\\University of Copenhagen, Blegdamsvej 17, 2100 Copenhagen, Denmark}

\date{\today}

\begin{abstract}
The large neutrino density in the deep interior of core-collapse supernovae and compact binary merger remnants makes neutrino flavor evolution non-linear because of the coherent forward scattering of neutrinos among themselves. Under the assumption of spherical symmetry, we model neutrino decoupling from matter in an idealized setup and present the first non-linear simulation of  flavor evolution in the presence of charged current and neutral current collisions, as well as neutrino advection. Within our framework, we find that flavor transformation occurs before neutrinos  fully decouple from matter, dynamically affecting the flavor distributions of all neutrino species and shifting the location of the neutrino decoupling surfaces. Our results call for further work as they may have implications on the explosion mechanism of supernovae, the nucleosynthesis of the heavy elements, as well as the observable neutrino signal, all of which is yet to be assessed.

\end{abstract}

\maketitle

\section{Introduction}

Neutrinos are fundamental particles in the physics of core-collapse supernovae~\cite{Burrows:2020qrp,Janka:2016fox}. Neutrinos are copiously emitted as matter accretes onto the proto-neutron star and transport energy and lepton number. According to the delayed neutrino-driven mechanism~\cite{1966ApJ...143..626C,1985nuas.conf..422W,Bethe:1985sox}, neutrinos are essential to revive the stalled shock wave, powering the explosion. Similarly, in neutrino-cooled disks stemming  from the merger of two neutron stars or  the collapse of rotating massive stars, neutrinos radiate most of the heat stemming from the turbulent flow,  the  neutrino pair-annihilation rate may contribute to power the jet of short gamma-ray bursts, and the neutrino-driven outflow launched from the disk  contributes, at least partially, to the synthesis of the heavy elements~\cite{Ruffert:1996by,Popham:1998ab,Beloborodov:2002af,Ruffert:1998qg,Just:2015dba,Cowan:2019pkx}.

Despite swift progress occurring in the last decade, 
neutrinos are treated as radiation in hydrodynamic simulations of core-collapse supernovae and compact binary merger remnants~\cite{1985S&T....70..231M,Mezzacappa:2020oyq,Shibata:2019wef,Foucart:2022bth,Fernandez:2022yyv}. However, they should be modeled through a quantum kinetic approach describing the evolution of the distributions of all six neutrino species (three neutrino flavors and their antiparticles), including flavor mixing~\cite{Sigl:1992fn,Vlasenko:2013fja,Vlasenko:2014bva,Serreau:2014cfa,Blaschke:2016xxt}. This approximation is adopted because solving the full quantum kinetic transport of neutrinos is a formidable task, entailing the solution of a $7$-dimensional non-linear problem (involving time, three spatial coordinates, energy, and two angular degrees of freedom), whose characteristic quantities rapidly vary by many orders of magnitude~\cite{Duan:2010bg,Mirizzi:2015eza,Tamborra:2020cul}. To date, this problem is not solvable with available computational resources.

Flavor evolution depends on the neutrino mixing parameters, as well as on the coherent forward scattering of neutrinos onto each other and on matter~\cite{Pantaleone:1992eq, Sigl:1992fn,Duan:2005cp,Duan:2006an,Duan:2006jv, Fogli:2007bk,Hannestad:2006nj}. Neutrino-neutrino interaction makes the equations of motion non-linear and  correlates the flavor evolution of neutrinos of different momenta. A full solution of the quantum neutrino transport problem does not exist yet.

The omission of neutrino conversion in hydrodynamical simulations was not deemed worrisome in supernovae as it was expected to occur beyond the shock radius,  and thus, not having an impact on the shock revival~\cite{Dasgupta:2011jf}. However, this picture has been shaken by recent developments~\cite{Sawyer:2005jk,Sawyer:2008zs,Sawyer:2015dsa,Chakraborty:2016lct,Izaguirre:2016gsx}, hinting that neutrino conversion could already develop in the proximity of the neutrino decoupling region~\cite{Tamborra:2017ubu, Abbar:2018shq, DelfanAzari:2019epo, Morinaga:2019wsv,DelfanAzari:2019tez, Nagakura:2019sig, Abbar:2019zoq, Glas:2019ijo, Capozzi:2020syn, Nagakura:2021hyb, Harada:2021ata}, with a  characteristic time scale that is proportional to the number density of neutrinos and antineutrinos (hence dubbed ``fast'' to distinguish it from the ``slower'' conversion driven by the vacuum frequency). The impact of  flavor conversion on the supernova physics remains to be understood.  Additionally,  in compact binary mergers, flavor conversion could affect the disk cooling rate as well as the neutron-to-proton ratio~\cite{Wu:2017qpc,Wu:2017drk,George:2020veu,Just:2022flt,Li:2021vqj}.

  In the context of fast flavor conversion, an important role is played by the angular distributions of (anti)neutrinos:  a crossing between the angular distributions of $\nu_e$ and $\bar\nu_e$, or a change of sign in the electron neutrino lepton number (ELN) within a certain angular range, constitutes a necessary condition to trigger a  flavor instability~\cite{Izaguirre:2016gsx,Morinaga:2021vmc,Dasgupta:2021gfs},  in the limit of vanishing mass-squared difference~\cite{Sawyer:2005jk, Sawyer:2015dsa, Tamborra:2020cul, Izaguirre:2016gsx, Shalgar:2020xns, Shalgar:2019qwg, Shalgar:2021wlj, Johns:2019izj, Chakraborty:2019wxe, Abbar:2018shq, Dasgupta:2016dbv}. The amount of induced flavor conversion is, however, unknown~\cite{Padilla-Gay:2021haz}.  
In the light of these developments, 
it becomes of paramount importance 
to ascertain whether the presence of ELN crossings is modified by flavor conversion and vice-versa. 

In the supernova core, for example, neutrinos  are trapped  with a mean-free path of $\mathcal{O}(10)$~m at typical baryon densities of $\mathcal{O}(10^{14})$~g/cm$^3$~\cite{Brown:1982cpw}. 
As the distance from the supernova increases and the baryon density decreases,  neutrinos start decoupling from matter. However, not all flavors of neutrinos decouple simultaneously due to the flavor dependent cross section of neutrinos. 
Their angular distributions are initially almost isotropic and, as the density decreases, become forward peaked and ELN crossings can form, also because of collisions and advection~\cite{Shalgar:2019kzy,Nagakura:2021hyb}. 
Recent work shows that, if flavor conversion should occur in the decoupling region, a non-trivial feedback between neutrino flavor transformations and collisions could take place~\cite{Shalgar:2019kzy,Shalgar:2020wcx,Hansen:2022xza,Sasaki:2021zld,Martin:2021xyl,Capozzi:2018clo,Johns:2021qby,Richers:2019grc}. 
Similarly, neutrino advection and the spatial dimensionality of the problem could also dynamically modify the conditions for flavor conversion~\cite{Shalgar:2019qwg,Richers:2021xtf}.

In this paper,  we present the first investigation of flavor conversion as the angular distributions of neutrinos evolve in the collisional regime within an idealized toy-model shell and in the presence of neutrino advection. 
Our work is organized as follows. We introduce the mean field equations of motion  in Sec.~\ref{sec1}, together with the modeling of the collision term. The decoupling of neutrinos in the absence and then in the presence of flavor conversion is discussed in Secs.~\ref{sec:no_conv} and \ref{sec2}, respectively. Section~\ref{sec3} focuses on the interplay between slow and fast conversion. The characteristic time scales of the problem are instead discussed in Sec.~\ref{sec:converg}. Finally, our findings are summarized in Sec.~\ref{sec:conclusions}. Appendix~\ref{appendix} provides additional material comparing  simulations with our standard resolution with others obtained with a smaller number of bins in radius.

\section{Mean field equations of neutrinos}
\label{sec1}
In this section, we introduce the kinetic equations of neutrinos. We also outline our  modeling of the collision term, which  mimics the hierarchy of decoupling of the different neutrino flavors, despite being heuristic. 
\subsection{Kinetic equations}
For the sake of simplicity, we  work in the two-flavor approximation, i.e.~in the $(\nu_e,\nu_x)$ basis~\cite{Shalgar:2021wlj,Capozzi:2020kge,Chakraborty:2019wxe,Capozzi:2022dtr}. The [anti]neutrino field is described through a $2\times2$ density matrix, $\rho(\vec{r}, \vec{p}, t)$ 
[$\bar\rho(\vec{r}, \vec{p}, t)$]. The diagonal elements of the density matrix ($\rho_{ii}$, with $i = e, x$) represent the occupation numbers of neutrinos of different species, while the off-diagonal terms ($\rho_{ij}$) encode flavor coherence. 
The evolution of the neutrino field at time $t$, location $\vec{r}$ and momentum $\vec{p}$  is governed by the following equation~\cite{Sigl:1993ctk}:
\begin{eqnarray}
\left(\frac{\partial}{\partial t} + \vec{v} \cdot \vec{\nabla}\right)\rho(\vec{r}, \vec{p}, t) = \mathcal{C} - i[H(\vec{r}, \vec{p}, t),\rho(\vec{r}, \vec{p}, t)]\ .
\label{mix:eom}
\end{eqnarray}
We work in spherical symmetry and assume only one energy mode $E$; hence,  $\rho(\vec{r}, \vec{p}, t) = \rho(r, E, \theta, t)$, with $\theta$ being the local polar angle defined with respect to the radial direction and  at a given $r$.
On the left-hand side of Eq.~\ref{mix:eom},  the advective term is,
\begin{eqnarray}
\vec{v}\cdot\vec{\nabla} = \cos\theta \frac{\partial}{\partial r} + \left(\frac{\sin^{2} \theta}{r}\right) \frac{\partial}{\partial \cos\theta}, 
\end{eqnarray}
in agreement with e.g.~Refs.~\cite{Rampp:2002bq,Nagakura:2022qko}.

The commutator on the right hand side of Eq.~\ref{mix:eom} encodes the flavor evolution of neutrinos.
The Hamiltonian, $H = H_{\textrm{vac}} + H_{\textrm{matt}} + H_{\nu\nu}$, governs the coherent evolution of neutrinos and is composed of the vacuum and neutrino self-interaction term: 
\begin{eqnarray}
H_{\textrm{vac}} &=& \frac{\omega}{2}
\begin{pmatrix}
-\cos 2 \vartheta_{\textrm{V}} & \sin 2 \vartheta_{\textrm{V}} \cr
\sin 2 \vartheta_{\textrm{V}} & \cos 2 \vartheta_{\textrm{V}} 
\end{pmatrix}\ , \\
H_{\nu\nu}&=&\mu_0 \int [\rho(\cos\theta^{\prime})-\bar{\rho}(\cos\theta^{\prime})] \nonumber\\
&\times& (1-\cos\theta \cos\theta^{\prime}) d\cos\theta^{\prime}\ ,
\end{eqnarray}
where we have absorbed the constant factor coming from the integration  over the azimuthal angle in $\mu_0$ and suppressed the dependence on $r$ and $t$ for the sake of brevity. The term, $H_{\textrm{matt}}$, linked to matter effects should also appear in $H$. However, the effect of the matter potential   is to suppress the vacuum mixing angle $\vartheta_{V}$~\cite{Esteban-Pretel:2008ovd}, hence we set $H_{\textrm{matt}} = 0$ and use   $\vartheta_{V}=10^{-3}$ instead~\footnote{A smaller $\vartheta_{V}$ would lead to a longer simulation time needed to achieve the quasi-steady state configuration. We have checked that the quasi-steady state configuration is virtually unaffected [for $\vartheta_{V}=10^{-8}$,  the quasi-steady state is reached at $10^{-4}$~s (as for the run with $\vartheta_{V}=10^{-3}$), and we find  a relative error of $1.7\%$ averaged over the radius with respect to the case with $\vartheta_{V}=10^{-3}$].}. 
The vacuum frequency is $\omega = \Delta m^2/2E$, with $ \Delta m^2$ being the squared mass difference.  For antineutrinos,  $\omega \rightarrow -\omega$ in $H_{\textrm{vac}}$, while all other terms of $H$ remain unchanged.

\subsection{The collision term}

The collision term in Eq.~\ref{mix:eom}, $\mathcal{C} \equiv \mathcal{C}(\vec{r}, E, t) =  \cemit + \cabsorb + \cdirch$,  includes  emission, absorption, and direction-changing collisions, respectively~\cite{weinberg_2019}. It encapsulates all the reactions of neutrinos with the matter background~\cite{Shalgar:2019kzy,1982ApJS...50..115B,OConnor:2014sgn,Mezzacappa:2020oyq}.
We assume that the collision term only affects the diagonal components of the density matrices~\footnote{We have tested that  the inclusion of the  off-diagonal components of the collision terms  following the method adopted in Ref.~\cite{Nagakura:2022qko} results in a negligible error on the final flavor configuration (the relative error averaged  over the radius is  $2.2\%$).}, as the time scales associated with flavor evolution are shorter than the collision ones~\cite{Tamborra:2020cul}.

\begin{figure}[t]
\includegraphics[width=0.45\textwidth]{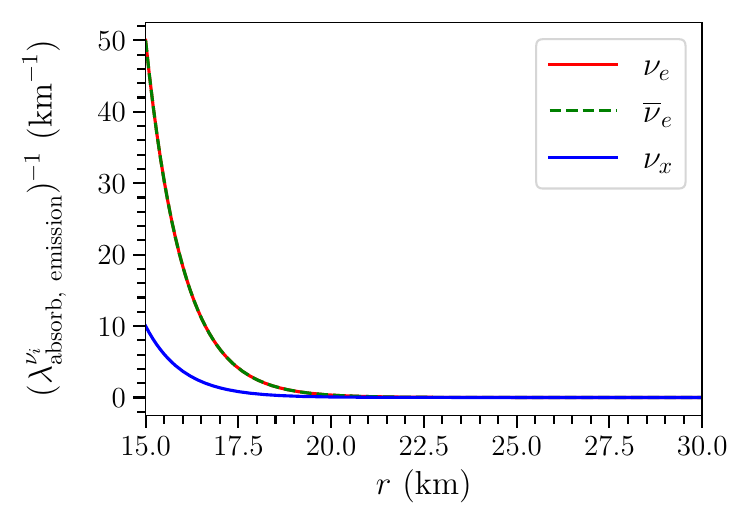}
\includegraphics[width=0.45\textwidth]{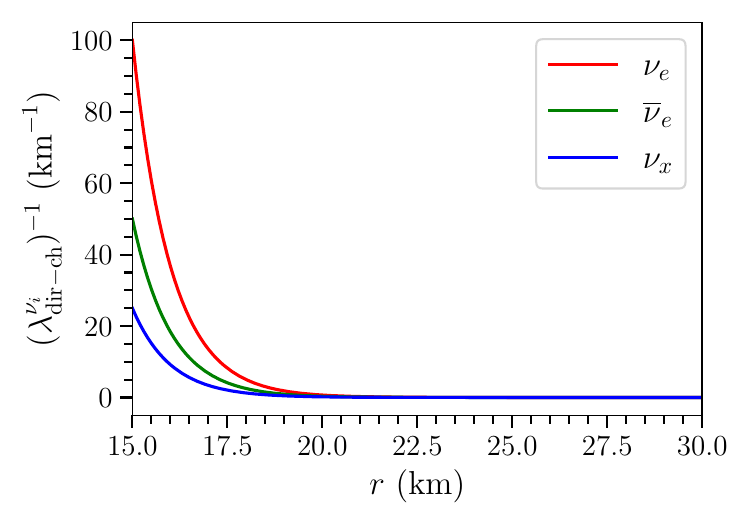}
\caption{Inverse length-scales associated with emission, absorption, and direction-changing scattering for $\nu_e$ (in red), $\bar\nu_e$ (in green), and $\nu_x$ (in blue) as functions of radius. The functional forms of these terms are reported in Table~\ref{Tab1}, see also Eqs.~\ref{cemission}--\ref{collform}.}
\label{MFPs}
\end{figure}

For the sake of simplicity, given that we intend to explore the interplay between the neutrino decoupling and the flavor conversion for the very first time, we rely on a heuristic collision term, that allows for neutrinos to transition from the trapping regime to free streaming within a small spatial range (i.e., $[15,30]$~km in our simulations, while providing decoupling regions and decoupling hierarchy among the different flavors inspired by Ref.~\cite{Tamborra:2017ubu}), since we are limited by the  size of our simulation shell---due to  technical complications induced by the treatment of flavor conversion.
For simplicity, we  retain the dependence of $\mathcal{C}$ on  $r$ and neglect its  angular dependence because it is dominated by the interaction of neutrinos with nucleons that are heavier than the neutrino average energies: 
\begin{eqnarray}
\label{cemission}
\cemit^{\nu_{e},\bar{\nu}_{e},\nu_{x},\bar\nu_x} &=& \frac{1}{\lambda_{\textrm{emission}}^{\nu_{e},\bar{\nu}_{e},\nu_{x},\bar\nu_x}(r)} \ , \\
\label{cabsorb}
\cabsorb^{\nu_{e},\bar{\nu}_{e},\nu_{x},\bar\nu_x} &=& - \frac{1}{\lambda_{\textrm{absorb}}^{\nu_{e},\bar{\nu}_{e},\nu_{x},\bar{\nu}_{x}}(r)} \rho_{ii}(\cos\theta)\ , \\
\label{cdirch}
\cdirch^{\nu_{e},\bar{\nu}_{e},\nu_{x},\bar\nu_x} &=& - \frac{2}{\lambda_{\textrm{dir-ch}}^{\nu_{e},\bar{\nu}_{e},\nu_{x},\bar\nu_x}(r)} \rho_{ii}(\cos\theta) \nonumber\\
&+& \int_{-1}^{1} \frac{1}{\lambda_{\textrm{dir-ch}}^{\nu_{e},\bar{\nu}_{e},\nu_{x},\bar\nu_x}(r)} \rho_{ii}(\cos\theta^{\prime}) d\cos\theta^{\prime}\ . 
\label{collform}
\end{eqnarray}
Each of the above equations refers to all flavors as denoted by the superscripts, and the flavor-dependent length-scales, $\lambda^{\nu_i}$,  are tabulated in Table~\ref{Tab1} and shown in Fig.~\ref{MFPs}. Equations~\ref{cemission} and \ref{cabsorb} determine the rate at which neutrinos traveling in the direction $\theta$ are created and absorbed, respectively. Equation~\ref{collform}  represents the probability that a neutrino  traveling along $\theta^{\prime}$  will change to the $\theta$ direction. While Eqs.~\ref{cemission} and \ref{cabsorb} do not conserve the number of neutrinos, Eq.~\ref{cdirch} preserves  the number of neutrinos.

In our simulations, neutrinos of all flavors are generated through collisions. The ratio between $\cemit$ and $\cabsorb$ determines $\rho_{ii}$ for all flavors at $r_{\textrm{min}}=15$~km. 
Since only $\cemit$ and $\cabsorb$ play a role in the trapping regime, the number density at $r_{\rm min}$ can be determined analytically in the steady state configuration. At larger $r$, $\cdirch$ also comes into play in shaping the angular distributions and no analytical solution exists. 
 In addition, in our simplified setup, we neglect Pauli blocking~\cite{1985ApJS...58..771B,Raffelt:1996wa}  and assume that the neutrino chemical potentials are negligible.

\begin{table}
\caption{Inverse length-scales associated with emission, absorption and direction-changing scattering for all the flavors (see Eqs.~\ref{cemission}--\ref{collform}). The function $\xi(r)=\exp(-(r-15)/\textrm{km})$ ensures that the collision term vanishes at large distances.}
\label{Tab1}
\begin{tabular}{|l|l|l|l|}
\hline
& $\nu_{e}$ & $\bar{\nu}_{e}$ & $\nu_{x}, \bar\nu_x$ \cr
\hline
\hline
$\lambda_{\textrm{emission}}^{\nu_i}$ (km) & 1/[50 $\xi(r)]$ & 1/[50 $\xi(r)]$ & 1/[10 $\xi(r)]$ \cr
\hline
$\lambda_{\textrm{absorb}}^{\nu_i}$ (km) & 1/[50 $\xi(r)]$ & 1/[50 $\xi(r)]$ & 1/[10 $\xi(r)]$ \cr 
\hline
$\lambda_{\textrm{dir-ch}}^{\nu_i}$ (km) & 1/[50 $\xi(r)]$ & 1/[25 $\xi(r)]$ & 1/[12.5 $\xi(r)]$ \cr
\hline
\end{tabular}
\end{table}

To  investigate the decoupling of neutrinos, we focus on the simulation shell from $15$ to $30$~km. 
 We use $7500 \times 150$ bins in $r$ and $\cos\theta$, respectively. Additional technical details on the numerical implementation and convergence are provided in Secs.~\ref{sec:no_conv} and \ref{sec:converg}, as well as Appendix~\ref{appendix} and Ref.~\cite{Shalgar:2022lvv}.
 
The radial profiles for the heuristic collision terms are shown in Fig.~\ref{MFPs} (see also Table~\ref{Tab1}).
We can see that $\lambda_{\rm{dir-ch}}^{-1}$ is the largest contribution for all flavors at small radii.
The direction changing scattering is dominated by neutral current interactions with nucleons.
Because of our single energy approximation, we  adopt an heuristic approach to model the collision term such that
$\nu_e$'s have the smallest  mean free path, while non-electron type neutrinos have the longest mean free path, as expected~\cite{1982ApJS...50..115B,OConnor:2014sgn,Mezzacappa:2020oyq}. Moreover, the fact that the $\bar{\nu}_{e}$ mean free path is larger than the $\nu_{e}$ one ensures that the $\bar{\nu}_{e}$'s decouple at a smaller radius than $\nu_{e}$'s~\cite{Keil:2002in, Raffelt:2003en}. 
As the radius increases, all inverse length scales decrease, until they are negligible for $r \gtrsim 25$~km, when neutrinos decouple from matter as shown in the top panel of Fig.~\ref{inidist}. 

\begin{figure}
\includegraphics[width=0.49\textwidth]{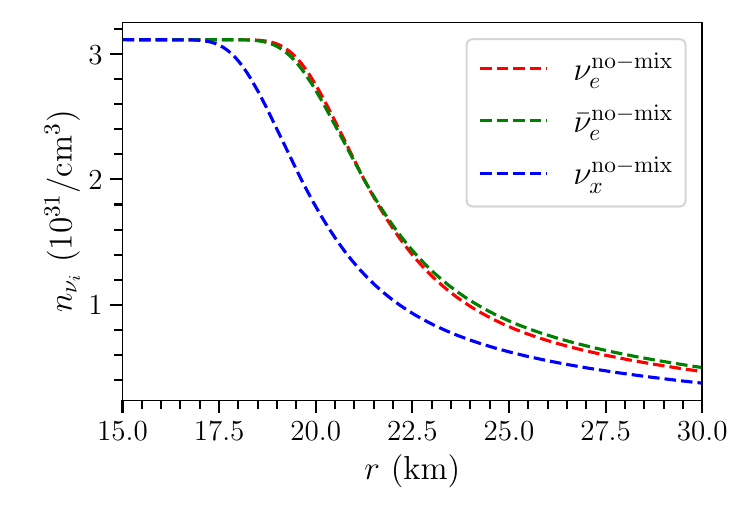}
\caption{Neutrino number densities of $\nu_e$ (in red), $\bar\nu_e$ (in green) and $\nu_x$ (in blue) as  functions of radius in the steady state configuration in the absence of neutrino mixing. The number density has no radial dependence in the trapping regime ($r \lesssim$ 18 km), while  it falls of like $1/r^{2}$ at larger radii ($r \gtrsim$ 25 km), as expected because the absorption and emission terms are very small.}
\label{numdenfig}
\end{figure}

As neutrinos start to decouple from matter,  the angular distribution for each neutrino flavor becomes forward peaked and the number density which is proportional to the diagonal components of the density matrix falls off.
This is clearly visible from  the radial profile of the number density for each flavor [which corresponds to $n_{\nu_i}({r}) = \mu_0 \int_{-1}^{1}\rho_{ii}(r,\cos\theta)d\cos\theta/(\sqrt{2} G_\textrm{F})$ in physical units] shown  in Fig.~\ref{numdenfig}, in the steady state configuration and in the absence of neutrino mixing. At radii larger than $\sim 25$ km, the emission and absorbtion terms are almost zero.
 In this limit we find that the neutrino number density falls off like $1/r^{2}$ as expected due to conservation of neutrino number for each flavor. Also, since there is almost no absorption or emission of neutrinos in the region of $r \gtrsim 25$ km, the angular distributions become  forward peaked as the only neutrinos that are present at these radii are the ones which have escaped the core and travel along directions that are close to the radial one.
 
 \section{Neutrino decoupling in the absence of  flavor conversion}
 \label{sec:no_conv}
For all the  simulations presented in this paper,  we use the central difference method for spatial derivatives, while an adaptive multi-step Adams-Bashforth-Moulton method is adopted for the temporal evolution with  absolute and relative tolerances of $10^{-12}$. The central difference method adopted to calculate the spatial derivative and the derivatives with respect to $\cos\theta$ cannot be applied to the edges of the simulation shell and hence to the related derivatives. As for  the derivative with respect to $\cos\theta$, we use the forward and backward difference method at $\cos\theta=-1$ and $\cos\theta=1$, respectively. For $r=r_{\textrm{min}}$, the boundary condition is set by requiring a classical steady state solution.
For $r=r_{\textrm{max}}$, there are two distinct domains that need to be considered independently. For $\cos\theta > 0$, neutrinos travel outwards and, therefore,  the boundary condition does not need to be specified. For $\cos\theta < 0$, the neutrino flux for all flavors is set to zero.

The kind of ELN crossing could affect the development of flavor conversion physics~\cite{Tamborra:2020cul,Nagakura:2019sig}. We choose a representative case in this work. However, we investigate the flavor conversion physics for a set of  different ELN crossings (i.e., different collision terms)  in Ref.~\cite{Shalgar:2022lvv} and show that our main  conclusions  are not qualitatively affected.

\begin{figure}
\includegraphics[width=0.49\textwidth]{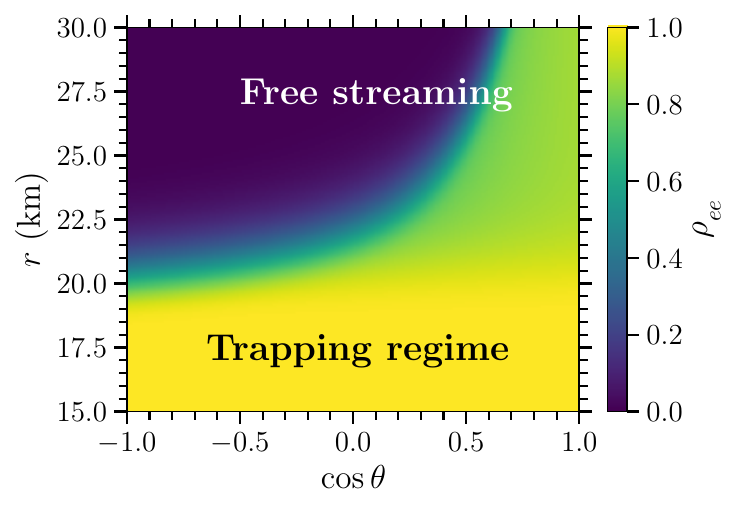}\\
\includegraphics[width=0.49\textwidth]{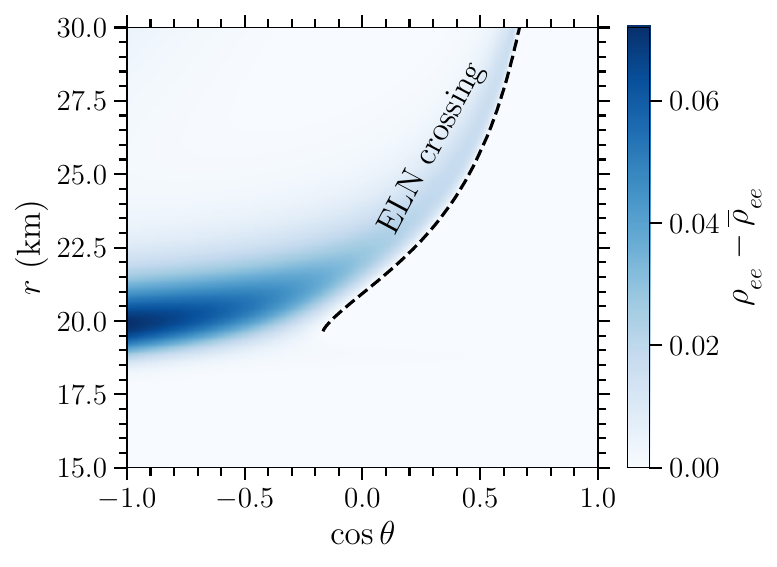}\\
\hspace{-1cm}\includegraphics[width=0.39\textwidth]{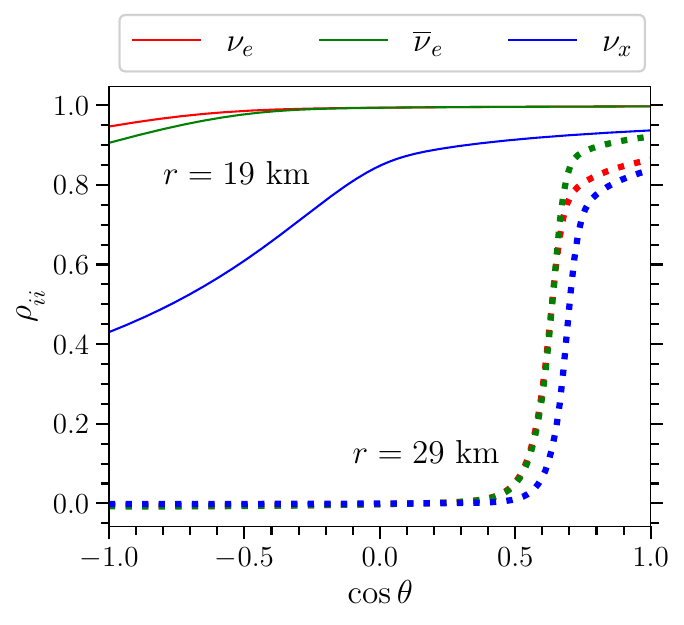}
\caption{Steady state neutrino flavor configuration in the absence of flavor mixing. These distributions (generated by imposing $H=0$ in Eq.~\ref{mix:eom} and extracted  at $t= 10^{-4}$~s) are the  ones adopted as input to investigate the effects of flavor conversion. {\it Top panel:} Contour plot of  $\rho_{ee}$ (proportional to the $\nu_e$ number density) in the plane spanned by $\cos\theta$ and $r$. 
{\it Middle panel:} Same as in the top panel but for $\rho_{ee}-\bar\rho_{ee}$ (proportional to the ELN density). The dashed line marks the region where ELN crossings develop.
{\it Bottom panel:} Angular distributions of $\nu_{e}$, $\bar{\nu}_{e}$ and $\nu_{x}$ at $r = 19$~km (solid) and $29$~km (dotted). As $r$ increases, the angular distributions   become  prominently forward peaked in a flavor-dependent fashion and ELN crossings develop. 
\label{inidist}}
\end{figure}

The flavor configuration displayed in Fig.~\ref{inidist} represents the classical steady state configuration obtained in the absence of neutrino flavor transformation.  Within our setup, the neutrino angular distributions are populated through  collisions and advection across the simulation shell. While we rely on heuristic collision terms and focus on an idealized simulation shell,  the decoupling regions and the hierarchy among the different flavors qualitatively reproduce the ones obtained from hydrodynamical simulations, e.g.~Ref.~\cite{Tamborra:2017ubu}, while also developing ELN crossings.

The decoupling of neutrinos from matter is driven by the flavor-dependent collision term along with neutrino advection~\cite{Shalgar:2019kzy,Ott:2008jb}. 
Solving Eq.~\ref{mix:eom} for $H = 0$ (i.e.~in the absence of flavor conversion) and no neutrinos in the simulation shell at $t=0$,  the collision terms in Eqs.~\ref{cemission}--\ref{cdirch} lead to a steady state configuration for each neutrino species after  a sufficiently large time interval  ($t \gtrsim 10^{-4}$~s). The top panel of Fig.~\ref{inidist} shows a contour plot of  $\rho_{ee}$  in the plane spanned by $\cos\theta$ and $r$. At small radii, the angular distribution ($15 \lesssim r \lesssim 18$~km) is uniform in $\cos\theta$; as $r$ increases, the angular distribution becomes gradually forward peaked and its backward part  ($\cos\theta \lesssim 0$) is progressively emptied as full decoupling is reached. 
A similar behavior holds for all (anti)neutrino species.

\begin{figure}[b]
\includegraphics[width=0.49\textwidth]{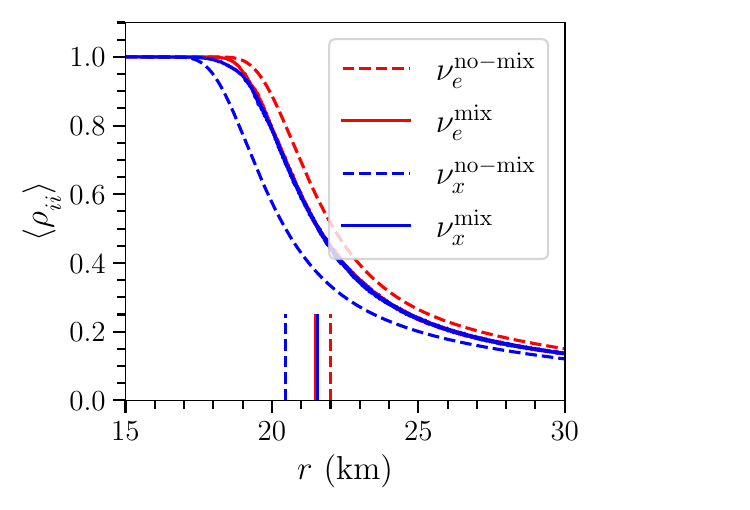}
\includegraphics[width=0.49\textwidth]{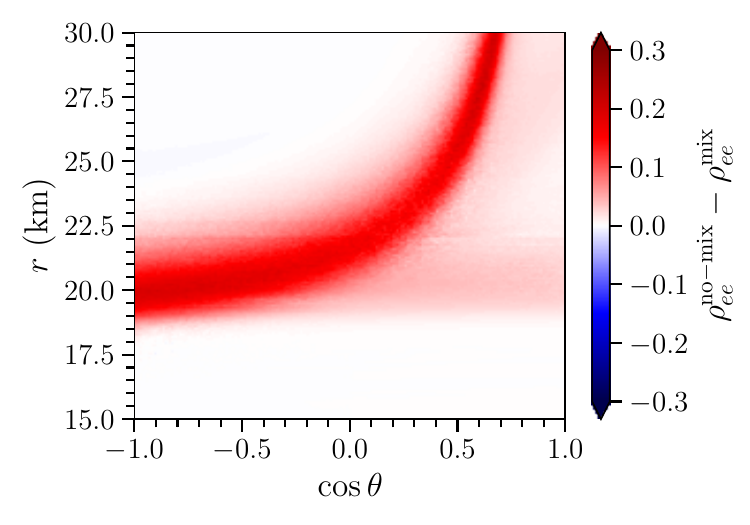}
\caption{Steady state neutrino flavor configuration in the presence of flavor mixing (extracted $5 \times 10^{-5}$~s after the classical steady state configuration is reached in our simulation).  {\it Top:} Angle averaged neutrino occupation numbers of $\nu_e$ (in red) and $\nu_x$ (in blue)  with (solid lines) and without (dashed lines) neutrino mixing as  functions of the radius.
The vertical lines mark the radii of decoupling  (approximately defined as the radius at which $\mathcal{F}_{\nu_i}  = 1/3$).
A similar trend holds for antineutrinos; however, note that flavor conversion induces a difference between $\nu_x$ and $\bar\nu_x$. {\it Bottom:} Contour plot of the difference between the $\nu_{e}$ occupation number without (when the classical steady state configuration is achieved) and with neutrino mixing in the plane spanned by $\cos\theta$ and $r$.  
Due to the collective nature of the neutrino flavor evolution and flavor lepton number conservation, the corresponding heatmap for
antineutrinos looks very similar and is not shown.} 
\label{angleave}
\end{figure}

As the presence of ELN crossings is crucial to the development of fast flavor conversion, the middle panel of Fig.~\ref{inidist} displays a contour of $\rho_{ee}-\bar\rho_{ee}$ in the plane spanned by $\cos\theta$ and $r$. ELN crossings are not prominent at small $r$, but they develop at $r \gtrsim   18$~km and affect different angular regions as $r$ increases. The dark blue band  to the left of the dashed line in Fig.~\ref{inidist} is the region with an excess of $\nu_{e}$ over $\bar{\nu}_{e}$ due to the fact that the $\bar{\nu}_{e}$ decoupling radius is smaller than the  $\nu_{e}$ one.
 The bottom panel of Fig.~\ref{inidist} displays snapshots of the  angular distributions of $\nu_{e}$, $\bar{\nu}_{e}$, and $\nu_{x} = \bar\nu_x$ at two representative radii. As $\mathcal{C}$ evolves as a function of $r$, the angular distribution of each flavor becomes forward peaked and ELN crossings  are visible. 

The dashed lines in the the top panel of Fig.~\ref{angleave} show the radial evolution of the angle-averaged occupation number for $\nu_e$ and $\nu_x$. In order to determine the decoupling surface, we  compute the
 flux factor: 
\begin{eqnarray}
\mathcal{F}_{\nu_i} = \frac{\int_{-1}^{1} \rho_{ii}(\cos\theta) \cos\theta d\cos\theta}{\int_{-1}^{1} \rho_{ii}(\cos\theta)  d\cos\theta}\ .
\label{ff}
\end{eqnarray}
and approximately define the radius of decoupling as the radius at which the flux factor is $\mathcal{F}_{\nu_i}  = 1/3$~\cite{Wu:2017drk,Tamborra:2017ubu}. The corresponding decoupling surfaces of $\nu_e$ and $\nu_x$ are  shown in the top panel of Fig.~\ref{angleave}. 

\section{Neutrino decoupling in the presence of  flavor conversion}
\label{sec2}

Switching on flavor mixing in Eq.~\ref{mix:eom} (i.e., $H \neq 0$), we assume that neutrinos have $\omega=0.32$~km$^{-1}$, which corresponds to  $E=20$ MeV for  $\Delta m^2 = 2.5\times 10^{-3}$ eV$^{2}$, and $\mu_0 = 10^4$~km$^{-1}$.  We adopt the steady state flavor configuration in Fig.~\ref{inidist} as the initial state. 

Figure~\ref{angleave} shows fascinating insights into the effect of flavor conversion on  neutrino decoupling. 
The solid lines in the top panel of Fig.~\ref{angleave} show the angle averaged occupation numbers of $\nu_e$ and $\nu_x$. With respect to the case without flavor conversion (dashed lines in Fig.~\ref{angleave}), we see that flavor conversion pushes the distributions of $\nu_e$ and $\nu_x$ towards each other, sensibly modifying them with respect to the case without flavor conversion. However, we stress that flavor equipartition is not a general outcome,  but it is linked to the specific flavor setup adopted in this paper; we have found other flavor configurations that do not lead to flavor equipartition (results not shown here), see also Refs.~\cite{Wu:2021uvt,Shalgar:2022lvv}. The impact of neutrino flavor evolution is  evident in the shift of the radius at which the angle-averaged $\rho_{ee}$ and $\rho_{xx}$ start falling, with consequences on the  decoupling radius.

One of the consequences of neutrino flavor conversion is a shift in the radius of neutrino decoupling. 
The top panel of Fig.~\ref{angleave} shows the flavor dependent decoupling surfaces with (solid) and without (dashed) flavor conversion for comparison. 
It is also important to note that neutrinos start to change their flavor at radii much smaller than their decoupling surfaces; this is in stark contrast with the approximation of flavor-independent neutrinosphere commonly adopted in the literature~\cite{Duan:2005cp,Duan:2006an,Duan:2006jv,Fogli:2007bk, Duan:2010bg,Mirizzi:2015eza}, 
under the naive expectation that  flavor evolution near the neutrinosphere would not be present due to collisional damping~\cite{Stodolsky:1974hm,Harris:1980zi,Stodolsky:1986dx}.

\begin{figure}[]
\includegraphics[width=0.49\textwidth]{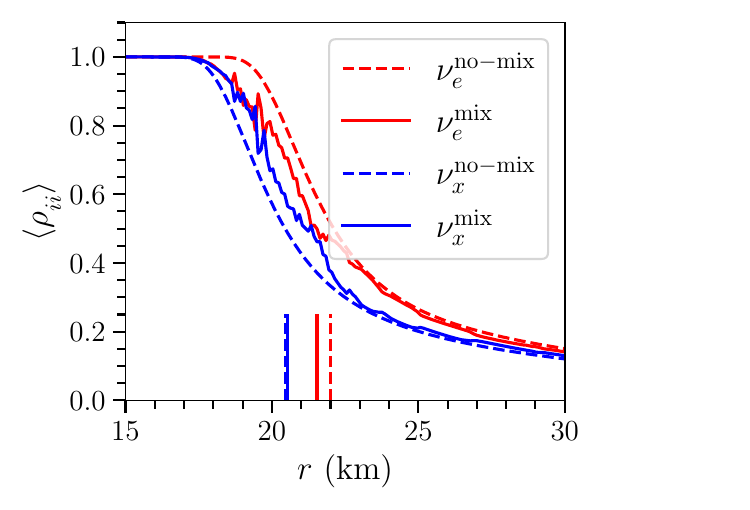}
\includegraphics[width=0.49\textwidth]{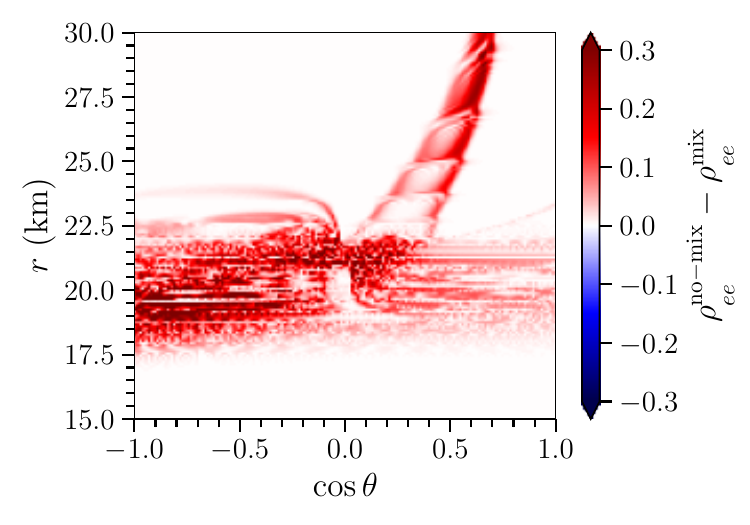}
\caption{Final flavor configuration obtained by  switching on flavor mixing, after the classical steady state is reached, and switching off collisions and neutrino advection (i.e., $H \neq 0$, $\mathcal{C}=0$, and $\vec{v} = 0$ in Eq.~1). The top and bottom panels are the analogous of the ones in Fig.~2, which instead take into account the dynamical effects on flavor mixing due to collisions and advection (i.e., $H \neq 0$, $\mathcal{C} \neq 0$, and $\vec{v} \neq 0$ in Eq.~1). }
\label{angintn}
\end{figure}

The bottom panel of Fig.~\ref{angleave} represents the difference in the $\nu_e$ occupation number in the absence and presence of neutrino mixing. As $r$ increases, a region with a deficit of $\nu_{e}$ forms due to flavor conversion (red band). 
This physics cannot be captured by any calculation that does not include advection and collision, together with flavor transformation--see Fig.~\ref{angintn} and \href{https://sid.erda.dk/share_redirect/fmRhOTrzOr/index.html}{the animations} for comparison. 

After obtaining a steady state configuration for all flavors (see Fig.~\ref{angleave}), we switch on flavor conversion and explore its interplay with neutrino advection and collisions. In order to highlight the dynamical effects  of neutrino advection and collisions on flavor conversion, we show in Fig.~\ref{angintn} the correspondent flavor outcome obtained by switching off collisions and advection once the steady state flavor configuration is achieved, i.e.~by letting flavor transformation operate in the absence of collisions and advection.

From Fig.~\ref{angintn}, one can see that collisions and neutrino advection not only affect the final angle-integrated flavor configuration, but also allow to spread the flavor instabilities across angular regions and spatial ranges~\cite{Hansen:2022xza,Shalgar:2020wcx,Shalgar:2019qwg}.
It is also worth noticing that the absence of neutrino advection makes the evolution of neutrino flavor much more oscillatory in nature.

\section{Interplay between slow and fast flavor instabilities}
\label{sec3}

Interestingly, despite the presence of ELN crossings and because of the shape of our ELN angular distributions (see middle panel of Fig.~\ref{inidist}) only a small radial range centered on $r \simeq 21$~km is affected by flavor instabilities in the limit of $\omega=0$ (i.e., when only fast conversion occurs). Because of  $\omega \neq 0$ and the system setup, slow collective neutrino transformations are also triggered at high densities, in contrast to what is commonly assumed in the literature.

In this section, we  choose to carry out the linear stability analysis for the homogeneous mode only,  as an illustrative example. However, the numerical simulations automatically take into account all modes. Hence, the interplay between slow and fast conversion is accounted for all modes consistently in the numerical simulations. The investigation of the interplay between fast and slow instabilities for non-homogeneous modes would require the development of a novel analytical framework because of the importance of the scattering terms  in our system setup; this task is left to future dedicated work. 

The linear stability analysis  consists of linearizing the equations of motion and  calculating the growth rate of the flavor instability~\cite{Banerjee:2011fj}.  By setting the collision and the advective terms to zero ($\vec{v}=0$), expanding Eq.~\ref{mix:eom} in the off-diagonal component,  and ignoring  $\mathcal{O}(\rho_{ex}^{2}$) and higher order terms, the solutions  are of the form:
\begin{eqnarray}
\rho_{ex}(\cos\theta,t) &\sim& \exp(-i\Omega t) \rho_{ex}(\cos\theta,0) \\
\bar{\rho}_{ex}(\cos\theta,t) &\sim& \exp(-i\Omega t) \bar{\rho}_{ex}(\cos\theta,0)\ .
\end{eqnarray}
The collective nature of  flavor evolution ensures that the eigenvalue $\Omega$ is the same for neutrinos and antineutrinos and also the same for all values of $\cos\theta$. When a flavor instability is present,  $\Omega$ has a positive imaginary component, called  growth rate and denoted by $\kappa$. It should be noted that, although the complex eigenvalues are always present in complex conjugate pairs in the absence of the collision term, this  is not the case in the presence of collisions~\cite{Cirigliano:2017hmk}, which we do not consider here.
However,  the presence of a neutrino flavor instability does not necessarily imply significant flavor transformation~\cite{Padilla-Gay:2020uxa}.

\begin{figure}[]
\includegraphics[width=0.49\textwidth]{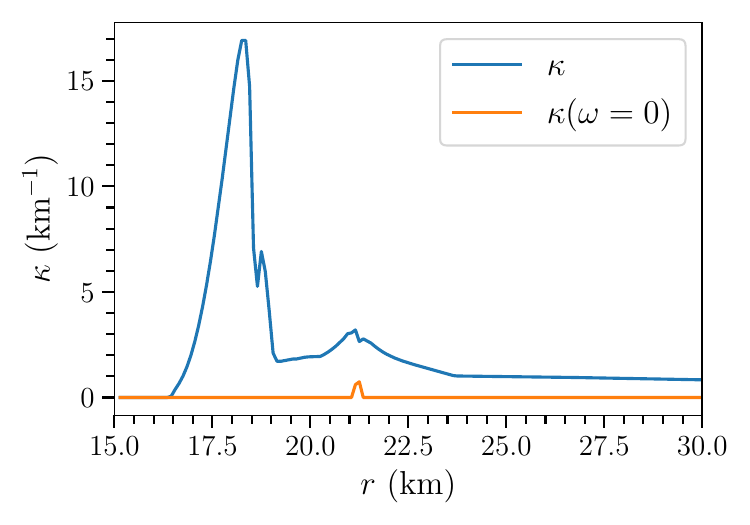}
\caption{Growth rate of the flavor instability as a function of  radius obtained by adopting the steady state flavor distributions. The orange line shows the growth rate for $\omega = 0$ (fast conversion), whereas the blue line shows the growth for $\omega =0.32$~km$^{-1}$ (slow conversion). Contrary to expectations, flavor mixing is induced by slow instabilities; fast flavor instabilities are only present  around $\sim 21$~km. }
\label{growth}
\end{figure}

Figure~\ref{growth} shows the growth rate as a function of radius for $\omega = 0$ (fast conversion)  and $\omega =0.32$~km$^{-1}$ adopted in our work.
Interestingly,  for most of the radial range, the neutrino flavor instability does not exist for $\omega=0$ for the homogeneous mode because of the shape of our ELN distributions. This implies that initially the neutrino flavor evolution is not driven by ELN crossings, but by the vacuum term for $k=0$.
Fast flavor instabilities are only present in a small radial range around $\sim 21$~km. In addition, once triggered, fast flavor mixing is further affected by the vacuum frequency~\cite{Shalgar:2020xns}. We stress that these findings imply an interplay between fast and slow conversion and do not intend to suggest that the slow modes are dominant for any $k$.
By comparing Fig.~\ref{growth} with Figs.~\ref{angleave} and \ref{angintn}, we note that although the neutrino flavor instability exists for $r \gtrsim 18$~km, the magnitude of the neutrino flavor transformation is not directly correlated to  $\kappa$, in agreement with the findings of Ref.~\cite{Padilla-Gay:2020uxa} in the context of fast flavor transformation.

\section{Characteristic scales of the problem}
\label{sec:converg}

In the context of fast conversion, the time scale of flavor evolution can be as large as $\mu^{-1}$ in  special cases (e.g., for $\omega=0$ and in the absence of collisions and advection). For more realistic cases  the time scales are smaller by a few orders of magnitude. 

One might naively confuse the time scale with the length scale over which spatial structure is present and in turn assume that this would infer the spatial resolution required to perform a reliable numerical study. However, as demonstrated in Ref.~\cite{Padilla-Gay:2020uxa}, we stress that spatial and temporal scales are not identical when advection is taken into account; it is possible to get reliable results by coarse-graining over the small spatial structure, if the latter is at all present, and  reproduce the average flavor ratio, see also Appendix~\ref{appendix} and Ref.~\cite{Padilla-Gay:2020uxa}. 
 The typical length scale of our problem should not be  considered to  be $\mathcal{O}(\mu_{0}^{-1})$, as in e.g.~Refs.~\cite{Richers:2021xtf, Richers:2021nbx, Richers:2022bkd,Wu:2021uvt,Bhattacharyya:2020jpj}---these works rely on a different system setup, with structures at length scales of  $\mathcal{O}(\mu_{0}^{-1})$ and periodic boundary conditions that may lead to cascades to smaller  length scale with time~\cite{Johns:2020qsk}.

\begin{figure*}
\includegraphics[width=0.39\textwidth]{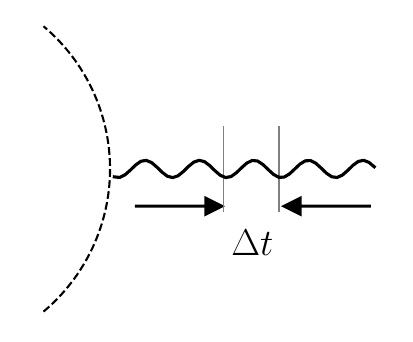}
\includegraphics[width=0.39\textwidth]{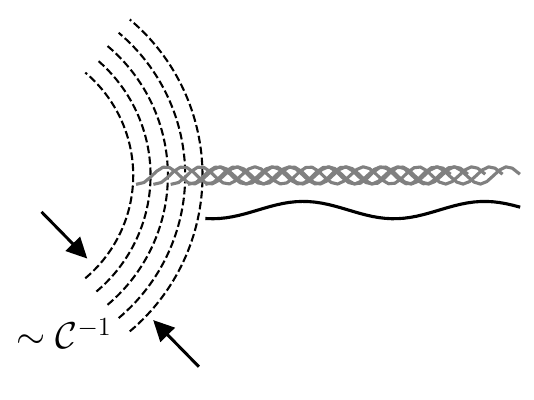}
\caption{Schematic representation of neutrinos being emitted by a decoupling region with infinitesimal extent (left panel) and neutrinos emitted from  an extended decoupling region composed of multiple emitting surfaces, each separated from the former and the following one by $\mathcal{O}(\mathcal{C}^{-1})$. For a density profile that is rapidly decreasing,  the decoupling region has a width that  roughly scales as  $\sim \mathcal{C}^{-1}$ (right panel).   In the first case, the typical length scale of the neutrino oscillation pattern is given by the product of the flavor conversion  time scale $\Delta t$ and the advection velocity. When an extended decoupling region is considered, the resultant oscillation pattern (black curve) is given by the superposition of multiple oscillation patterns (gray curves),  each separated from the other ones by $\mathcal{O}(\mathcal{C}^{-1})$.}
\label{cartoon}
\end{figure*} 
In our case, the characteristic length scale of the problem   is determined by the flavor conversion one, as well as the collision and advection  scales. If  neutrinos were assumed to be emitted by a single surface, as sketched in the left panel of Fig.~\ref{cartoon}, the typical length scale of the problem would be given by the product of the flavor conversion  time scale and the speed of light (i.e., the advection velocity). This would be the length scale  seen in a setup similar to the neutrino-bulb model, where the decoupling region has an infinitesimal extent. However, our decoupling region is much more extended  than the oscillation length scale,  see right panel of Fig.~\ref{cartoon}. Hence,  at a given point, the resultant neutrino field is given  by the  neutrinos emitted from various locations in the decoupling region {(i.e.~the small-scale structures that one would see when considering one single emitting surface are smeared as a result of the superposition of oscillating patterns overlapping with each other, as sketched in the right panel of Fig.~\ref{cartoon})}. 

For a density profile that is rapidly decreasing, as ours,  the extent of the decoupling region is roughly determined by the inverse of the collision term, $\sim \mathcal{C}^{-1}$.
Although it is not possible to estimate a priori the exact length scale associated with the neutrino field in the presence of flavor transformation, it is clear that the resultant length scale should be between the oscillation  and the collision length scales. 
In summary, $\mu_{0}^{-1}$ constitutes the minimum length scale expected for flavor conversion, the maximum being the collision length scale. 
As a consequence spatial resolution of  $\mathcal{O}(\mu_{0}^{-1})$ may not be needed for numerical convergence because of the interplay of flavor conversion with collisions and advection.

\section{Outlook}
\label{sec:conclusions}

For the first time, we follow the flavor evolution while neutrinos decouple in an idealized shell, in the presence of collisions and advection. We find that neutrino flavor conversion occurs well before all flavors decouple from the matter background and the neutrino decoupling surfaces are affected by flavor transformation. 

While dealing with flavor transformation, advection and collisions simultaneously, the collision term adopted in this work relies on heuristic functions that allow us to transition from uniform to forward peaked neutrino distributions within a small spatial region because of the technical challenges induced by the treatment of flavor conversion in the stellar core. A further exacerbation of our findings could be determined by the inclusion of all six flavors,  the energy dependence of the neutrino distributions, and the azimuthal emission angle of neutrinos~\cite{Shalgar:2020xns,Shalgar:2021wlj,Shalgar:2021oko,Capozzi:2020kge,Richers:2021xtf,Capozzi:2022dtr}.   
Additional effects on the decoupling physics may arise from relaxing the assumption of spatial homogeneity~\cite{Duan:2014gfa}. Not only can spatial inhomogeneity lead to additional flavor instabilities, neutrino advection causes dynamical effects which can only manifest in an inhomogeneous system~\cite{Shalgar:2019qwg}.

Despite its caveats and the idealized simulation setup, this work clearly demonstrates that the neutrino decoupling physics is affected by flavor transformation, with possible implications on the neutrino energy deposition in the supernova gain layer and the physics of compact merger remnants that remain to be assessed. 
The fact that the dynamics of neutrino decoupling changes due to neutrino flavor transformation implies that the neutrino properties at the radius of decoupling may be different from  the ones usually computed in hydrodynamic simulations treating neutrinos as radiation. However, a robust assessment of the radiated
neutrino spectra in the presence of flavor conversion will have to take into account the 
hydrodynamic feedback on the thermodynamic properties and a more realistic implementation of the collision term. The occurrence of flavor conversion  in the vicinity of the flavor-dependent neutrinospheres puts in perspective the existing literature naively assuming flavor-independent neutrinospheres and a larger radius for the onset of flavor conversion.

\acknowledgments

We  thank Eve Armstrong, Rasmus S.~L.~Hansen, Christopher Rackauckas, and Anna Suliga for useful discussions, as well as Thomas Janka and Georg Raffelt for insightful  feedback on the manuscript.
We are grateful to the Villum Foundation (Project No.~13164), the Danmarks Frie Forskningsfonds (Project No.~8049-00038B),  the MERAC Foundation, and the Deutsche Forschungsgemeinschaft through Sonderforschungbereich SFB~1258 ``Neutrinos and Dark Matter in Astro- and Particle Physics'' (NDM).

\appendix
\section{Comparison with simulations with smaller spatial resolution}
\label{appendix}
\begin{figure}
\includegraphics[width=0.49\textwidth]{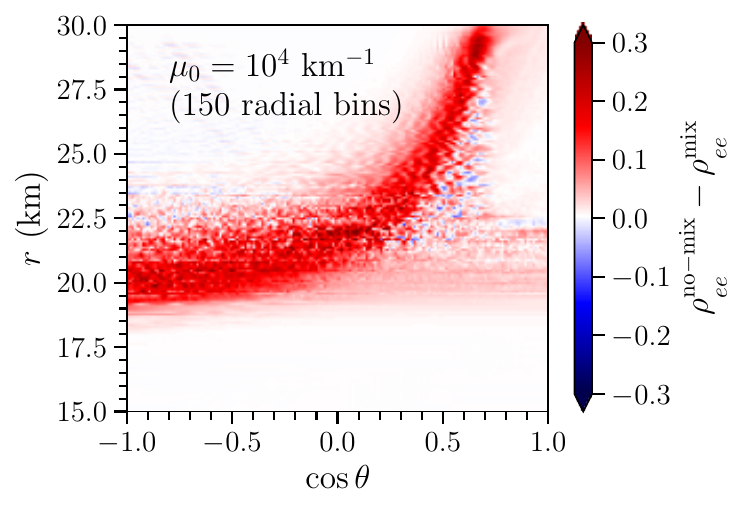}
\includegraphics[width=0.49\textwidth]{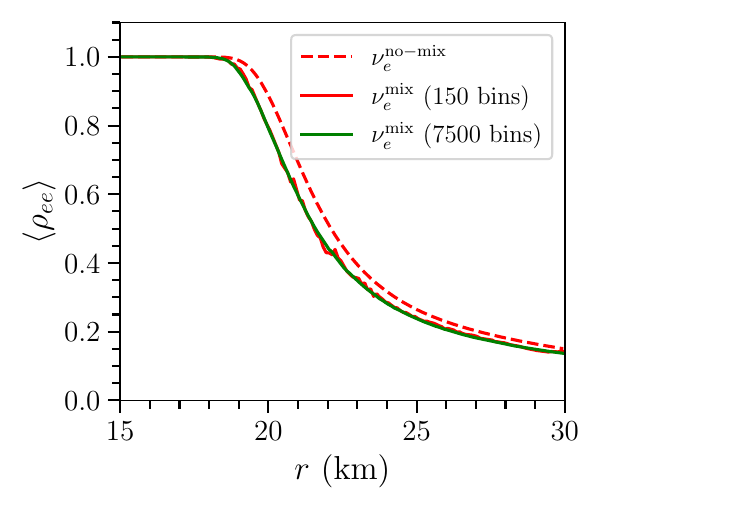}
\includegraphics[width=0.49\textwidth]{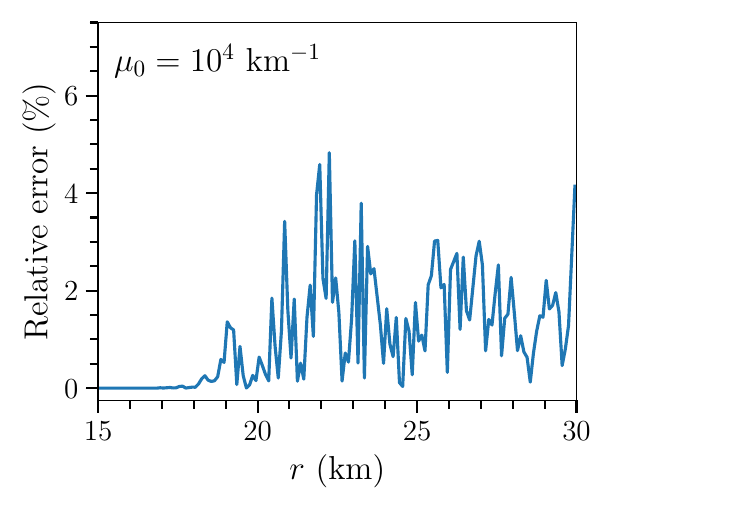}
\caption{Comparison between simulations with a different number of radial bins for $\mu_0=10^{4}$ km$^{-1}$. The top panel shows the difference between $\rho_{ee}$ with and without mixing for $\mu_0=10^{4}$~km$^{-1}$ computed using $150$ spatial bins. The middle panel shows the initial angle averaged $\rho_{ee}$ as red dashed line. The red solid line and green solid line shows are the angle averaged $\rho_{ee}$ calculated using $150$  and $7500$  spatial bins. The red solid line is difficult to see due to the overlap with the green solid line. The bottom panel shows the relative error between the two simulations as a function of $r$. The overall error averaged over radius is $1.1\%$.}
\label{FigS51}
\end{figure}

\begin{figure*}
\includegraphics[width=0.49\textwidth]{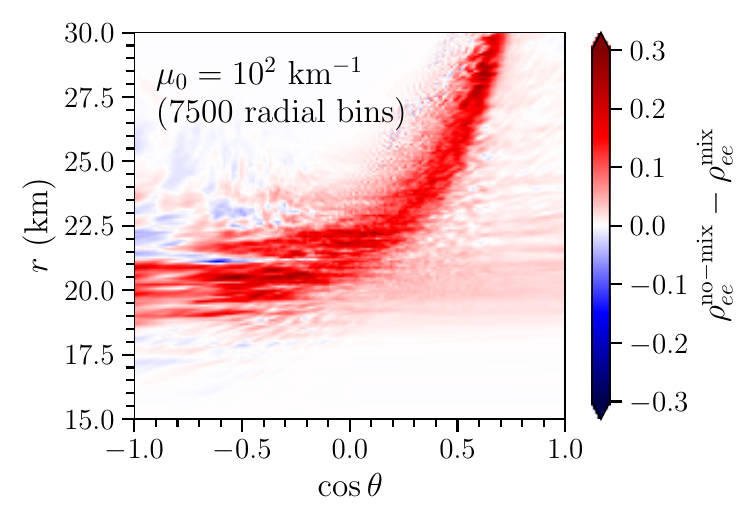}
\includegraphics[width=0.49\textwidth]{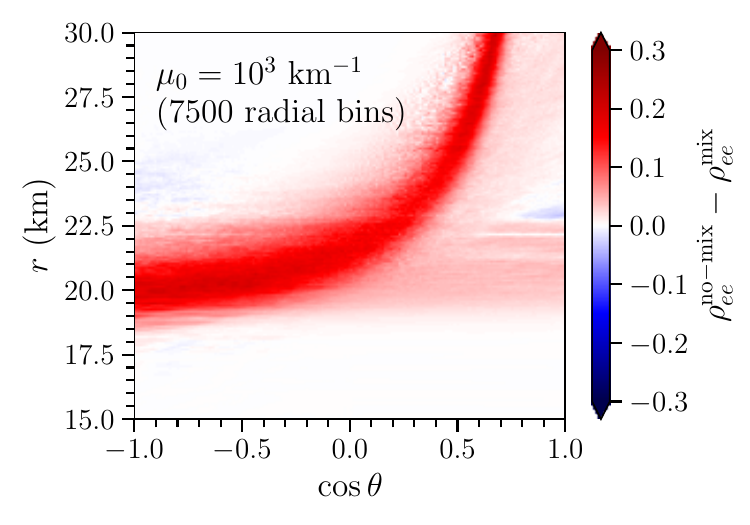}
\includegraphics[width=0.49\textwidth]{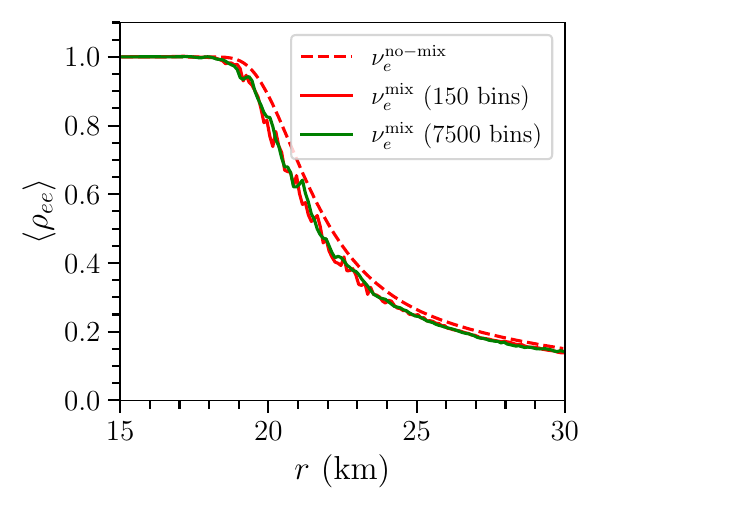}
\includegraphics[width=0.49\textwidth]{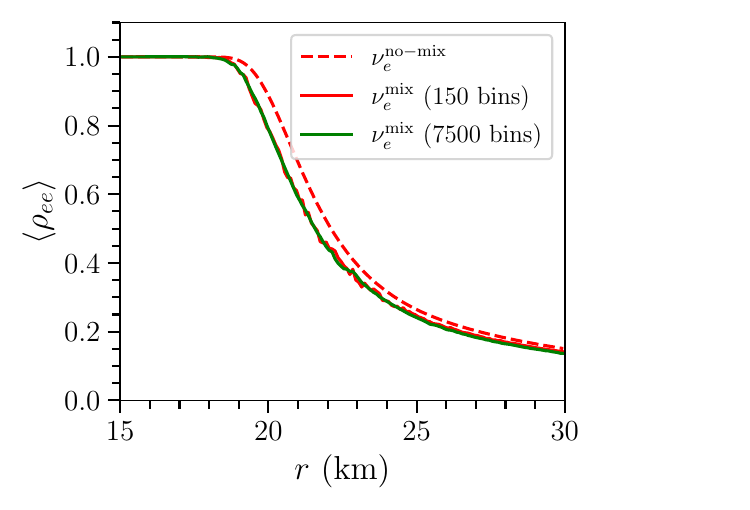}
\includegraphics[width=0.49\textwidth]{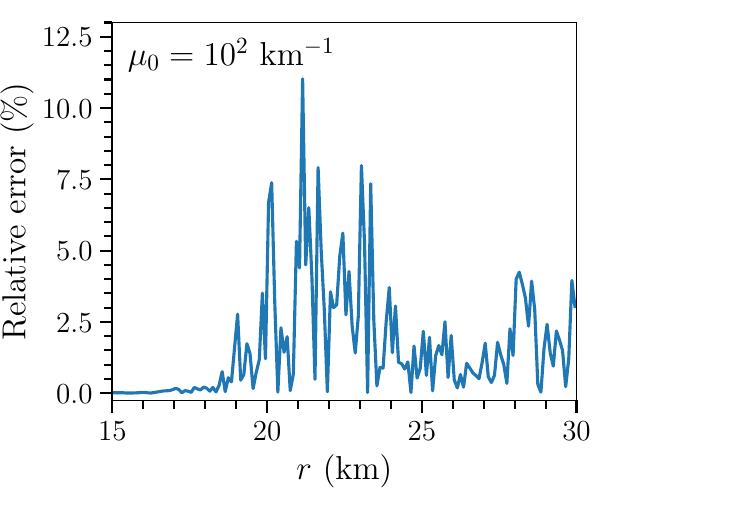}
\includegraphics[width=0.49\textwidth]{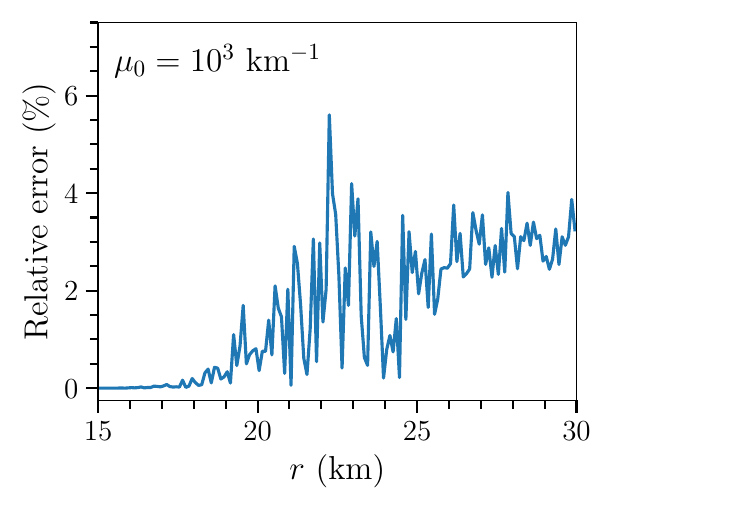}

\caption{Same as Fig.~\ref{FigS51} but with $\mu_0=10^{2}$~km$^{-1}$ (left) and $\mu_0=10^{3}$~km$^{-1}$ (right).}
\label{FigS52}
\end{figure*}

In this appendix, we perform a simulation with a smaller number of spatial bins ($150$) with respect to the one used in the main text.
In order to do so, we consider our benchmark  system and perform simulations with the self-interaction strength at $r_{\rm{min}}$ being $\mu_{0}=10^{4}$, $10^{3}$, and $10^2$~km$^{-1}$ with $150$ spatial bins and compare the results with simulations performed with $7500$ spatial bins adopted in the main text (i.e., $50$ times higher spatial resolution). The choice of a smaller $\mu_0$ allows us to perform numerical simulations with a spatial resolution that is accurate enough  to resolve length scales of the order of $\mu_{0}^{-1}$ or smaller.

Figure~\ref{FigS51} shows results  for  $\mu_0=10^{4}$ km$^{-1}$ for our  two simulations  with  $150$  and  $7500$ radial bins (same as the one presented in the main text), with all other simulation inputs unchanged with respect to  our benchmark model (including  the number of angular bins kept fixed to $150$).
 The top panel shows the difference in the electron neutrino number density, with and and without flavor conversion obtained using $150$ spatial bins. Despite the appearance of small scale structure due to the coarser graining in radius,the results are same as the ones presented in Fig.~\ref{angleave} within reasonable numerical errors. This is evident from  the middle panel of Fig.~\ref{FigS51}, where we show the comparison between the number densities averaged over angle for the simulations with $150$ and $7500$ spatial bins.  
 The bottom panel of Fig.~\ref{FigS51} displays the  relative error between the two simulations. The error averaged over the radial range is $1.1\%$. This  shows that  $150$ spatial bins are sufficient to capture the main features of the  flavor evolution, despite the appearance of small scale structures.

\begin{figure*}[!th]
\includegraphics[width=0.49\textwidth]{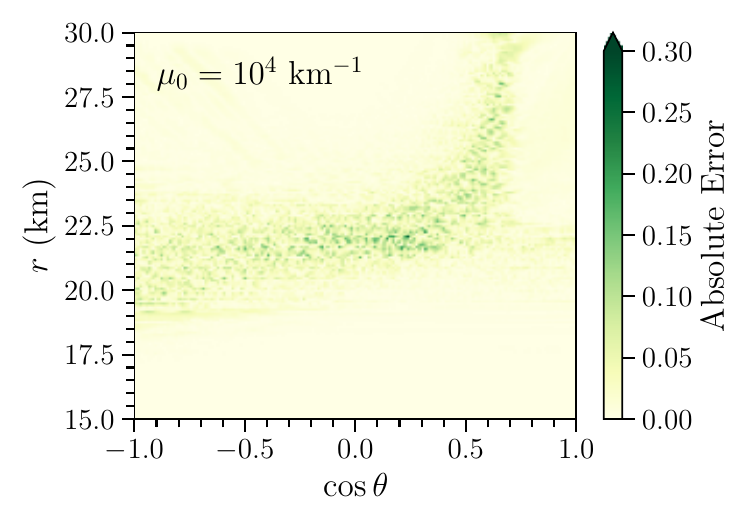}
\includegraphics[width=0.49\textwidth]{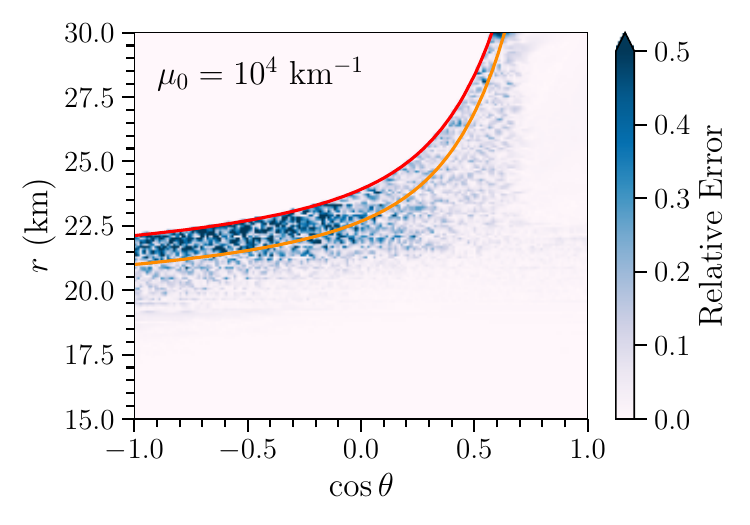}
\caption{{\it Left: } Absolute error in $\rho_{ee}^{\mathrm{no-mix}}-\rho_{ee}^\mathrm{mix}$ in the plane spanned by  $\cos\theta$ and $r$, and for $\mu_{0}=10^4$~km$^{-1}$. The error has been obtained by comparing the coarse grained results of the simulation with 7500 spatial bins with the results of the simulation with 150 spatial bins. {\it Right: } Relative error in $\rho_{ee}^\mathrm{mix}$ for regions where $\rho_{ee}^\mathrm{no-mix}$ is greater than $0.1$. For reference, the loci of points where $\rho_{ee}^\mathrm{no-mix}$ is equal to $0.1$ and $0.5$ are shown in red and orange, respectively. 
}
\label{errheatmaps}
\end{figure*}

Figure~\ref{FigS52} shows  a  comparison between simulations using smaller values of $\mu_{0}$ ($100$ and $1000$~km$^{-1}$, left and right panels, respectively). From the bottom panels, we can see that the average relative error is $1.66$ and $1.63\%$ for $\mu_{0} = 100$ and $1000$~km$^{-1}$, respectively. 

In all cases, the average relative error between the simulations with $150$  and $7500$ radial bins  can reach  up to $4$--$5\%$ in a few radial bins but on average is less than $1$--$2\%$. 
The relative error has been computed  by coarse-graining the results with $7500$ radial bins and averaging over batches of $50$ radial bins, to compare the results with the ones from the numerical simulation that used $150$ radial bins. Since a quasi-steady state configuration is achieved, the relative errors quoted above overestimate the actual error due small time-dependent fluctuations that are present at every given point when the simulation is stopped. 

It is easy to see that small time-dependent fluctuations are  the main source of the error by investigating  the error without averaging over the angle. Figure~\ref{errheatmaps} shows the absolute error in $\rho_{ee}^\mathrm{mix} - \rho_{ee}^{\mathrm{no-mix}}$ in the left panel and the relative error  in the right panel for the simulations with $\mu_0 =10^4$~km$^{-1}$  (see also Figs.~\ref{angleave} and \ref{FigS51}).

\begin{figure}
\includegraphics[width=0.49\textwidth]{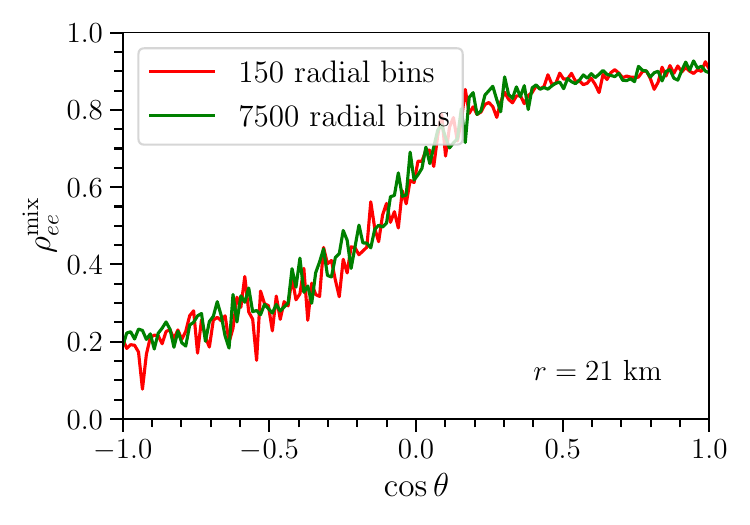}
\includegraphics[width=0.49\textwidth]{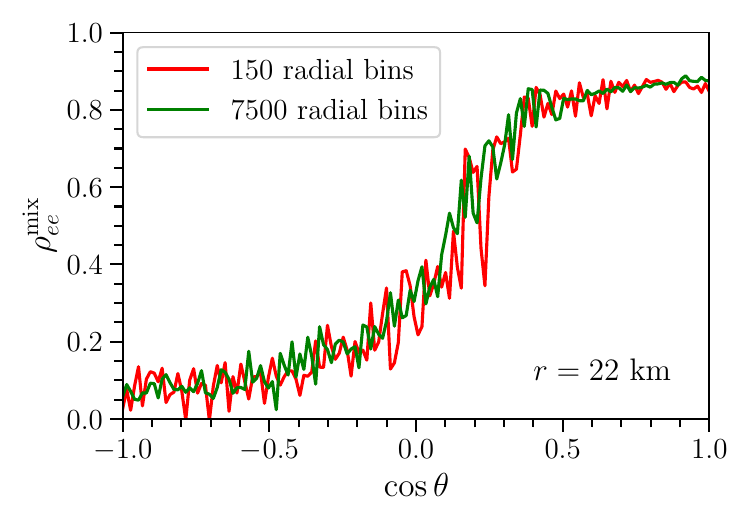}
\caption{ Angular distributions of $\nu_e$ after flavor conversion, extracted at  $r=21$ and $r=22$~km and for $\mu_{0}=10^{4}$ km$^{-1}$ for the simulations with $150$  (red) and $7500$ (green) radial bins. The two curves are in very good agreement. 
}
\label{angdisterr}
\end{figure}

The absolute error  in the left panel of Fig.~\ref{errheatmaps} provides an accurate representation of the error in our simulation, which   is about $0.1$--$0.15$ throughout the simulation domain.
 The relative error is shown in the right panel of Fig.~\ref{errheatmaps} for completeness. 
We plot the relative error only for the regions where $\rho_{ee}^\mathrm{no-mix} \gtrsim 0.1$ (see red contour in the right panel of Fig.~\ref{errheatmaps}).
To guide the eye, we  also show an orange contour where $\rho_{ee}^\mathrm{no-mix} \simeq 0.5$. The right panel shows that the relative error between the low- and high-resolution simulations is less than $5$--$10\%$. However, it can reach up to $40\%$ for $\cos\theta < 0$  where the relative error is not an informative quantity since there are very few neutrinos in the tail of the angular distribution. 
 { We stress that the relative error is  ill-defined in such regions    and should be considered with extreme caution. 
 This becomes clear by looking at  Fig.~\ref{angdisterr} where we show the $\nu_e$ angular distribution when the quasi-steady state is reached for  $r=21$ and $r=22$~km with $\mu_{0}=10^{4}$ km$^{-1}$. We plot $\rho_{ee}^{\textrm{mix}}$ obtained from the simulation with $150$ spatial bins, alongside with the correspondent angular distributions obtained from the simulation with $7500$ spatial bins. The angular distributions obtained for the simulations  with $150$ and $7500$ bins are in very good agreement, yet if one were to compute the relative error e.g.~for $\cos\theta = -0.9$, one would obtain an error above  $40\%$; this clearly shows the misleading information provided by the relative error in  the right panel of Fig.~\ref{errheatmaps} for some regions of the parameter space spanned by $\cos\theta$ and $r$.
}

\begin{figure}
\includegraphics[width=0.49\textwidth]{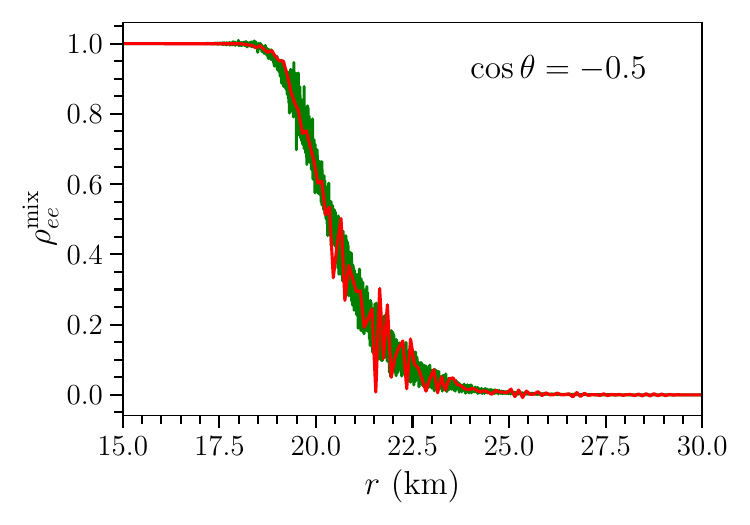}
\includegraphics[width=0.49\textwidth]{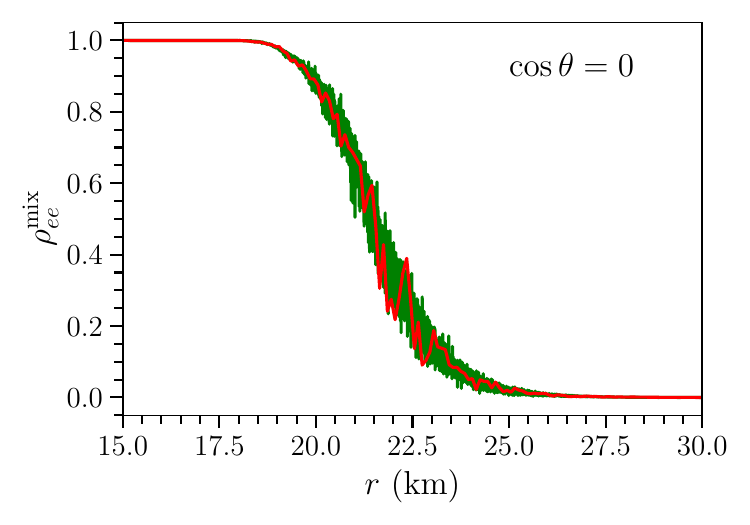}
\includegraphics[width=0.49\textwidth]{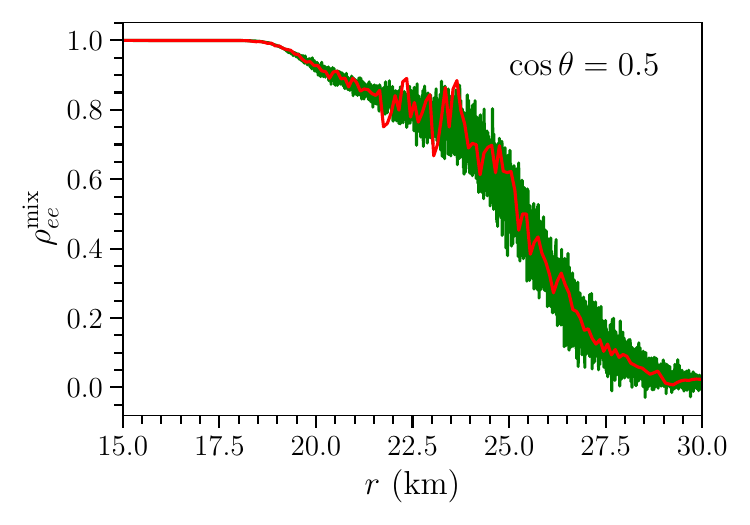}
\caption{Radial profile of  $\rho_{ee}$ after flavor conversion, extracted at  $\cos\theta=-0.5$, $0$, and $0.5$ (from top to bottom, respectively) for $\mu_{0}=10^{4}$ km$^{-1}$ for the simulations with $150$  (red) and $7500$ (green) spatial bins. 
}
\label{rad_evol}
\end{figure}

Figure~\ref{rad_evol} shows the radial evolution of $\rho_{ee}^{\textrm{mix}}$ obtained from the simulations with $150$ and  $7500$ spatial bins for   $\cos\theta=-0.5$, $0$, and $0.5$ (from top to bottom panels, respectively). One can clearly see that the low-resolution curve tracks the average behavior of the high-resolution one.

It is worth highlighting  that the value of $\mu_0$ reported above  is the self-interaction strength  at $r_{\textrm{min}}$; as neutrinos start decoupling, the effective strength of  self-interaction decreases as the fourth power of the radius, which is automatically taken into account in our simulations.



\bibliography{full-letter.bib}

\end{document}